\documentclass[12pt]{article}
\usepackage[latin1]{inputenc}
\usepackage{graphicx,epsfig}
\usepackage{amsmath}
\usepackage{amsfonts}
\usepackage{amssymb}
\usepackage{amsthm}
\usepackage{bm}
\usepackage{slashed}
\usepackage{color}
\def\be{\begin{equation}}
\def\ee{\end{equation}}
\def\ba{\begin{eqnarray}}
\def\ea{\end{eqnarray}}

\begin{document}

\thispagestyle{empty}
\begin{center}

\vspace{0.4cm}
{\Large {\bf Accelerated detectors in Dirac vacuum: \\ \vspace{0.5cm} the effects of horizon fluctuations }}\\
\vspace{1.0cm}
{\large {\bf C. H. G. B\'essa \footnote{{e-mail: chgbessa@cbpf.br}}},
{\bf J. G. Due\~nas\footnote{e-mail: jgduenas@cbpf.br} }, {\bf N. F. Svaiter\footnote{e-mail: nfuxsvai@cbpf.br} }} \\
\vspace{0.2cm}
{\it Centro Brasileiro de Pesquisas F\'{\i}sicas, Rua Dr. Xavier Sigaud 150, 22290-180, Rio de Janeiro, RJ, Brazil.} \\
\vspace{1.0cm}

\end{center}

\begin{abstract}
We consider an Unruh-DeWitt detector interacting with a massless Dirac field. Assuming that the detector is moving along an hyperbolic trajectory, we modeled the effects of fluctuations in the event horizon using a Dirac equation with random coefficients. First, we develop the perturbation theory for the fermionic field in a random media. Further we evaluate corrections due to the randomness in the response function associated to different model detectors.

\end{abstract}

\noindent{ {\it Keywords:} \\ Unruh-DeWitt detector \\ Horizon fluctuations \\Disordered medium \\Quantum gravity effects}

\section{Introduction}

In a previous paper \cite{akms11}, the spectral density associated to a scalar field assuming event horizon fluctuations was presented. In particular, it was considered a two-level uniform accelerated system interacting with a massless scalar field. Assuming the two-level system prepared in the ground state and the field in the Minkowski vacuum state, the main result of the paper was to show how the transition rates are modified by event horizon fluctuations. For modeling the event horizon fluctuation it was assumed that the scalar field satisfies a massless Klein-Gordon equation with random coefficient. It was found a modified response function with the same local temperature found in the non-fluctuating case. The correction due to the event horizon fluctuations has a Fermi-Dirac factor. A similar result found previously by Takagi \cite{t86} in a quite different situation. For a careful discussion for the apparent inversion of statistics see for example the Ref. \cite{unruh86}.

The main motivations for such kind of studies follows the ideas present in Ref. \cite{kms10}. In this work it was proposed in condensed matter physics an analog model for quantum gravity effects which in turn is based on results obtained by Ford and collaborators \cite{pe9,pe10} and Hu and Shiokawa \cite{pe11}. Two general features of waves propagating in random fluids were considered. First, acoustic perturbation in a fluid
define discontinuity surfaces that provide a causal structure with sound cones. Second, propagation of acoustic excitations in random media are generally described by wave equations with random speed of sound \cite{pe13,pe14,pe15,pe16}. A quantum scalar field theory associated to acoustic waves was analyzed in a situation where the velocity of propagation of the acoustic wave and consequently the sound cone fluctuates. In this approach a stochastic ensemble of fluctuating geometries was assumed.

In this paper we follow the ideas present in Ref. \cite{akms11}. Basically we study how a particular model for the fluctuations of a black-hole event horizon can affect the transition rate of a two-level system \cite{unruhdet, dewittdet}. Let us consider the line element of a four-dimensional Schwarzschild space-time describing a non-rotating uncharged black-hole of mass $M$:

\begin{equation}
ds^2=\biggl(1-\frac{2M}{r}\biggr)dt^2-\biggl(1-\frac{2M}{r}\biggr)^{-1}dr^2-r^{2}d\Omega^{\,2},
\label{det1}
\end{equation}
where $d\Omega^{\,2}$ is the metric of a unit $2$-sphere.
Close to the horizon $r\approx 2M$, the line-element given by Eq. (\ref{det1}) can be written as

\begin{equation}
ds^{2}=\biggl(\frac{\rho}{4M}\biggr)^2\,dt^{2}-d\rho^{2}-4Md\Omega^{\,2},
\label{det2}
\end{equation}
where $\rho(r) = \sqrt{8M(r-2M)}$ and the quantity $4Md\Omega^{\,2}$ describes the line element of a  $2$-sphere of radius $4M$. In turn, the other contribution can be identified with the line element of the two-dimensional Rindler edge by setting $t = 4Ma\tau$ and $\rho = \frac{e^{a\xi}}{a}$, for $0 < \rho < \infty$ and $-\infty < t < \infty$.

We can summarize our approach saying that one model to take into account the effects of black hole event horizons fluctuations is the following: we assume that event horizon fluctuations are modeled basically by a centered, stationary and Gaussian random processes. Under these conditions, we obtain a random differential equation for the Dirac field that cannot be solve exactly. Therefore, we implement a perturbation theory, similar to the one developed in Ref. \cite{pe21}, in our case associated to a massless Dirac field.

The organization of the paper is as follows: In section \ref{section2} we discuss the perturbation theory for the fermionic field in a disordered medium. In section \ref{section3} we present the modified positive Wightman function due to the fluctuating horizon. In section \ref{section4} we study the modifications to transition probabilities of a Unruh-DeWitt detector due the event horizon fluctuations. Finally, section \ref{section5} contains our conclusions. To simplify the calculations we assume the units to be such that $\hbar = c = k_{B} = 1$.

\section{Dirac field in disordered medium}\label{section2}

We begin by writing a random Dirac equation

\begin{equation}\label{eq1}
\left[i\gamma^0(1+\mu({\bf x}))\frac{\partial}{\partial t} - i\gamma^i\nabla_i - m\right]\Psi(t, {\bf x}) = 0.
\end{equation}
Here, $\Psi(t, {\bf x})$ represents a fermionic field,  $\gamma^0, \gamma^i$ are the Dirac matrices and $\mu({\bf x})$ is a dimensionless random function of the spatial coordinates. We consider a zero-mean random function, i.e.,

\begin{equation}\label{noise1}
\langle\mu({\bf x})\rangle_\mu = 0
\end{equation}
and for simplicity we suppose white-noise correlations, i.e.,

\begin{equation}\label{noise2}
\langle\mu({\bf x})\mu({\bf x'})\rangle_\mu = \sigma^2\delta({\bf x} - {\bf x'}).
\end{equation}
The symbol $\langle ...\rangle_\mu$ denote ensemble averaged of noise realizations and $\sigma$ represents the strength of the noise. We also suppose that the noise is Gaussian distributed. These last statements are pragmatic one. This is one of the simplest models one can choose which can exhibit light cone fluctuations. We will also see in next sections that this choice implies in a loss of stationarity in the Wightman function to the Rindler noise and the Wiener-Khintchine theorem is no longer valid \cite{Eber}.

To proceed, let us define the Fourier transforms to the Dirac field and also to the noise:

\begin{equation}
\Psi(t, {\bf x}) = \int \frac{d\omega}{2\pi} \frac{d{\bf k}}{(2\pi)^3}
e^{-i(\omega t -{\bf k}.{\bf x})}\Psi(\omega, {\bf k}),
\end{equation}

\begin{equation}
\mu({\bf x}) = \int \frac{d{\bf k}}{(2\pi)^3} e^{i{\bf k}.{\bf x}}\mu(\omega, {\bf k}).
\end{equation}

Similar to Ref. \cite{pe21} we can split Eq. \eqref{eq1} in a free and perturbed part $(L_0 + L_1)\Psi = 0$, where $L_0$ and $L_1$ are matrices in the coordinates or momentum spaces. We have

\begin{equation}
L_0 = i\gamma^0\frac{\partial}{\partial t} - i\gamma^i\nabla_i - m
\end{equation}
and
\begin{equation}
L_1 = i\gamma^0\mu({\bf x})\frac{\partial}{\partial t}.
\end{equation}
In the momentum space these operators are written as

\begin{equation}
\mathcal{L}_0 = \left[\gamma^0\omega + \gamma^ik_i - m\right]\delta({\bf k}- {\bf k'})
\end{equation}
and

\begin{equation}
\mathcal{L}_1 = \gamma^0\omega \mu({\bf k} - {\bf k'}).
\end{equation}

Thus, it is possible to define the full (operator valued) fermionic Green function, $S = (\mathcal{L}_0 + \mathcal{L}_1)^{-1}$. Assuming a weak noise, a perturbative expansion for $S$ is given by

\begin{equation}
S = S_0 - S_0\mathcal{L}_1S_0 + S_0\mathcal{L}_1S_0\mathcal{L}_1S_0 - ...
\label{eq:expansion1}
\end{equation}
where $S_0$ is just the inverse of the free operator and it is Fourier transformed by:

\begin{equation}
S_0(t, {\bf x}) = \frac{1}{(2\pi)^4}\int d^4k e^{-ik(x - x')}S_0(\omega, {\bf k}).
\end{equation}
In the massless case, $S_0(\omega, {\bf k})$ is given by the well known expression:

\begin{equation}
S_0(\omega,{\bf k}) = -\frac{\gamma_0\omega + \gamma^ik_i}{\omega^2 - {\bf k}^2}.
\end{equation}

After performing the average over the fluctuations, which can be seen also as an average in the ensemble of geometries in Eq. \eqref{eq:expansion1}, we can write down the connected two-point Green function related to the Dirac field as a Dyson equation

\begin{equation}
\langle S\rangle_{\mu}=S_0 - S_0\Sigma\langle S\rangle_{\mu},
\label{eq:Dyson-eq}
\end{equation}
where $\Sigma$ is the irreducible self-energy. The first contribution given by $\bar{S}^{(1)}(t,{\bf x})$ is of second order in random function $\mu$. In the Fourier representation it assumes the form given by:

\begin{eqnarray}
\bar{S}^{(1)}(t,{\bf x}) = \int \frac{d^4k}{(2\pi)^4}{\bar{S}}^{(1)}(\omega, {\bf k})e^{-ik(x - x')},
\label{eq:correction-coor}
\end{eqnarray}
where

\begin{equation}\label{eqs12}
{\bar{S}}^{(1)}(\omega, {\bf k}) \equiv \sigma^2\omega^2 \frac{\gamma^\mu k_\mu}{\omega^2 -
{\bf k}^2}\gamma^0\Sigma(\omega, {\bf k})
\gamma^0\frac{\gamma^\delta k_\delta}{\omega^2 - {\bf k}^2}.
\end{equation}
In the above equation the self-energy $\Sigma(\omega, {\bf k})$ is given by

\begin{equation}\label{self1}
\Sigma(\omega, {\bf k}) = \int d^3{\bf k'}\frac{(\gamma^0\omega - \gamma^i{k'}_i)}{\omega^2 - {\bf k'}^2}.
\end{equation}
Only the first part of the integral in Eq. \eqref{self1} contributes giving $2i\pi^2\omega^2\gamma^0$. A detailed calculation given in appendix A shows that the correction to the positive frequency Wightman function is given by

\begin{eqnarray}\label{eqfim1}
\bar{S}^{(1)}(t,{\bf x}) = \frac{\sigma^2}{2D^6}\gamma^0I_1 + \frac{\sigma^2}{2D^6}\left(\gamma^1 + \gamma^2 + \gamma^3\right)I_2
\end{eqnarray}
which

\begin{eqnarray}\label{eqfim2}
I_1 = -3684\Delta t^6 + 7539\Delta t^4\Delta{\bf x}^2 - 8328\Delta t^2 \Delta{\bf x}^4 - 1005\Delta{\bf x}^6\\
\nonumber +i\epsilon(18504\Delta t^5 - 24732\Delta t^3\Delta{\bf x}^2 + 9354\Delta t\Delta{\bf x}^4)
\end{eqnarray}
and

\begin{eqnarray}\label{eqfim3}
I_2 = -1660\Delta t^5\Delta{\bf x}
 - 2024\Delta t^3\Delta{\bf x}^3 - 156\Delta t\Delta{\bf x}^5 \\ \nonumber +
 i\epsilon(6620\Delta t^4\Delta{\bf x} +5056\Delta t^2\Delta{\bf x}^3 +
12\Delta{\bf x}^5).
\end{eqnarray}
We have used for $I_1$ and $I_2$ an first order expansion: $(x + a)^n \approx x^n + nx^{n - 1}a$ and $D$ has been defined by $D \equiv (\Delta t - i\epsilon)^2 - \Delta{\bf x}^2$.

We are now able to apply this result to the calculation of the response function of the Unruh-DeWitt detector. This will be done in next sections.

\section{Wightman functions and Rindler noise}\label{section3}

Before consider the Unruh-DeWitt detector we need to understand how the Wightman function and the Rindler noise changes if we take into account event horizons fluctuations. First, it is known that for an uniformly accelerated reference frame with proper acceleration $\alpha$, Rindler's coordinates are the appropriate coordinates to be used. The relation to the Minkowskian coordinates is given by

\begin{align}
\nonumber t&=\alpha^{-1}\sinh{\alpha\tau}\\
          x&=\alpha^{-1}\cosh{\alpha\tau}
      \label{eq: Rindler-coordiantes}
\end{align}
where $\tau$ is the observer's proper time. In the observer's proper reference frame it is necessary to Fermi-Walker transport the Dirac field due its non-stationarity \cite{t86}. We define a proper spinor in the following form

\begin{align}	
\hat{\Psi}(\tau) &= S_\tau\Psi(x(\tau)),
\label{eq:FW-fields}	
\end{align}
where the corresponding matrix $S_\tau$ takes care of the Fermi-Walker transport. Here the observer is, by definition, a Rindler observer, in a given time $\tau$ related to the laboratory frame we define the boosts $\Lambda_\tau$, which the non-vanishing components are given by

\begin{align}\label{eq:FW-fields2}	
\left(\Lambda_\tau\right)_0^{0} &= \left(\Lambda_\tau\right)_1^{1} = \cosh\alpha\tau, \nonumber\\
\left(\Lambda_\tau\right)_1^{0} &= \left(\Lambda_\tau\right)_0^{1} = -\sinh\alpha\tau, \nonumber\\
\left(\Lambda_\tau\right)_2^{2} &= \left(\Lambda_\tau\right)_{3}^{3} = 1.
\end{align}
So the corresponding matrix of the transformation of the spinor is given by

\begin{equation}\label{eq:FW-fields3}	
S_\tau = e^{\frac{1}{2}\alpha\tau\gamma_0\gamma_1} = \cosh(\alpha\tau/2) + \gamma_0\gamma_1\sinh(\alpha\tau/2).
\end{equation}

The Wightman function  will be also Fermi-Walker transported due its definition in terms of the Dirac Field. The ``proper" Wightman function in the accelerated reference frame is called the Rindler noise for the Dirac field and it is defined by:

\begin{equation}
\mathcal{S}(\tau, \tau') = S_\tau\langle S^{+}(x(\tau), x(\tau'))\rangle_{\mu} S_{-\tau'}.
\label{eq:Dirac-noise}
\end{equation}

Changing the variables $\xi = \tau-\tau'$ and $\eta=\tau+\tau'$ the total Green function is given by

\begin{equation}\label{eq:stotal}
\mathcal{S}(\xi, \eta) =  \mathcal{S}^{(0)}(\xi,\eta) + \mathcal{S}^{(1)}(\xi,\eta).
\end{equation}

The Rindler noise associated to the free case (non-random Dirac equations) is given by

\begin{equation}\label{freecase}
\mathcal{S}^{(0)}(\xi,\eta) = \frac{i}{2\pi^2}\left(\frac{\alpha}{2}\right)^3
\left[\gamma^0\cosh{\left(\frac{\alpha\eta}{2}\right)} + \gamma^1\sinh{\left(\frac{\alpha\eta}{2}\right)}
\right]\frac{S_{-\eta}\sinh(\frac{\alpha\xi}{2})}{\sinh^4(\frac{\alpha\xi}{2}-i\epsilon)},
\end{equation}
which is the result already known in literature. Using Eqs.\eqref{eqfim1}, \eqref{eqfim2} and \eqref{eqfim3} we find that the Rindler noise, taking into account fluctuations of the event horizon is given by:

\begin{align}\label{eq:correction_S}
 \mathcal{S}^{(1)}(\xi,\eta) =& -\left(\frac{\alpha}{2}\right)^6\left\{\sigma_{1}^2(\eta)\gamma^0S_{-\eta}
 \frac{\sinh(\frac{\alpha\xi}{2})}{\sinh^7(\frac{\alpha\xi}{2}
 - i\epsilon_{1})} \right. \\ \nonumber &\left. + \sigma_{2}^2(\eta)\gamma^1S_{-\eta}
 \frac{\sinh(\frac{\alpha\xi}{2})}{\sinh^7(\frac{\alpha\xi}{2}
 - i\epsilon_{2})}\right\},
\end{align}
where  some terms were absorbed by the $\epsilon_i$ factor. The strength noises were rewritten by $\sigma_{1}^2(\eta) = -\sigma^2F(\eta)$ and $\sigma_{2}^2(\eta) = -\sigma^2H(\eta)$ which are positive functions and grow with $\eta$, while the functions $F(\eta)$ and $H(\eta)$ are given by Eqs. (\ref{eq:fghj}) and (\ref{eq:fghj3}) in \ref{apendiceb}. This result shows how the fluctuating event horizon modifies the Rindler noise associated to the Dirac field. In the next section we will apply these results and the results found in previous section to find the response function associated to the Unruh-DeWitt detector.

\section{Unruh-DeWitt detectors for Dirac particles}\label{section4}

In this section we use our results obtained with the Dirac field to study the response and transition probabilities associated with the Unruh-DeWitt detector \cite{soffel} and \cite{iyer}. We assume an uniform accelerated detector coupled to a massless Dirac field $\Psi$ with the following interacting Hamiltonian

\begin{equation}
H_{int} = c_{1}m(\tau)\overline{\hat{\Psi}}(\tau)\hat{\Psi}(\tau).
\label{eq:lagrangian}
\end{equation}
In the Eq. (\ref{eq:lagrangian}) $x^{\mu}(\tau)$ is the world line of the two level system with respect to the proper time $\tau$,  $m(\tau)$ is the monopole moment operator
and $\overline{\Psi}$ denotes a Dirac conjugate to the field $\Psi$, and $c_{1}$ is the coupling constant between the detector and the Dirac field.

We consider the detector as a point-like object with a two energy levels structure given by $\omega_g < \omega_e$ and eigenstates $|g\rangle$ and
$|e\rangle$ respectively \cite{det1} and \cite{langlois}. The gap energy between both states would be $E = \omega_e - \omega_g$. Defining the initial state of the system at $\tau = 0$ as $|\tau_i\rangle = |g\rangle\otimes|\psi_i\rangle$ and a final state at time $\tau$ by $|\tau_f\rangle = |e\rangle\otimes|\psi_f\rangle$ we can apply perturbation theory to compute the probability of transition of the Unruh-DeWitt detector. Let us assume that the Dirac field is prepared in the Minkowski vacuum state $|0, M\rangle$. In first order perturbation theory, we get that the probability of transition $P(\tau, 0)$ is given by

\begin{equation}
P(\tau, 0) = c_1^2|\langle e|m(0)|g\rangle|^2R(E, \tau, 0),
\end{equation}
where the detector selectivity is given by $c_1^2|\langle e|m(0)|g\rangle|^2$.  Now we will only concentrate in the response function which is given by

\begin{align}
R(E, \tau, 0) = &\int_0^\tau d\tau'\int_0^\tau d\tau'' e^{-iE(\tau'-\tau'')} \\ \nonumber \times &\langle 0,M|\overline{\hat{\Psi}}(\tau')\hat{\Psi}(\tau')\overline{\hat{\Psi}}(\tau'')\hat{\Psi}(\tau'')|0, M\rangle.
\label{eqfunctionudw}
\end{align}
Then, in this case we have to calculate a four-point correlation function. Due to the Gaussian nature of the noise this four-point correlation function can be compute in terms of a product of two-point correlation functions. In this case, one can find \cite{langlois}:

\begin{equation}
\langle 0,M|\overline{\hat{\Psi}}(\tau')\hat{\Psi}(\tau')\overline{\hat{\Psi}}(\tau'')
\hat{\Psi}(\tau'')|0,M\rangle = Tr[(S^+(\tau', \tau''))^2],
\label{eq:eqtrace}
\end{equation}
where $Tr$ is the trace and $S^+(\tau',\tau'') = \langle 0,M|\hat{\Psi}(\tau')\overline{\hat{\Psi}}(\tau'')|0,M\rangle$ is the positive frequency Wightman function. Introducing again the variables $\xi = \tau' - \tau''$ and $\eta = \tau' + \tau''$ the response function may be written as \cite{nb}

\begin{equation}
R(E,\tau)=\frac{1}{2}\int_{-\tau}^{\tau}d\xi\int_{|\xi|}^{2\tau-|\xi|}d\eta e^{-iE\xi}Tr[(S^{+}(\xi,\eta))^2].
 \label{eq:response-function2}
\end{equation}

After averaging over the fluctuations, we can introduce  Eq. \eqref{eq:stotal} into Eq. \eqref{eq:response-function2} to get the one-loop correction to the response function. Hence, we expand the squared Wightman function in Eq. \eqref{eq:eqtrace} resulting in four contributions to the response function, where omitting the arguments of the Wightman functions it can be expressed by:

\begin{equation}
Tr((\mathcal{S}^{+})^2)=Tr(\mathcal{S}_0^2 + \mathcal{S}_0\mathcal{S}^{(1)}
+\mathcal{S}^{(1)}\mathcal{S}_0 + (\mathcal{S}^{(1)})^2).
 \label{eq:trace-exp}
\end{equation}

Substituting the Eq. (\ref{eq:trace-exp}) in Eq. (\ref{eq:response-function2}) we can write $R(E,\tau)=R_{0}(E,\tau)+R_{1}(E,\tau)$, where

\begin{equation}
R_{0}(E,\tau)=\frac{1}{2}\int_{-\tau}^{\tau}d\xi\int_{|\xi|}^{2\tau-|\xi|}d\eta e^{-iE\xi}Tr(\mathcal{S}_0^2) .
 \label{eq:response-function3}
\end{equation}
and

\begin{equation}
R_{1}(E,\tau)=\frac{1}{2}\int_{-\tau}^{\tau}d\xi\int_{|\xi|}^{2\tau-|\xi|}d\eta e^{-iE\xi}Tr(\mathcal{S}_0\mathcal{S}^{(1)}
+\mathcal{S}^{(1)}\mathcal{S}_0).
 \label{eq:response-function4}
\end{equation}

In the asymptotic limit the first term $R_{0}(E,\tau)$ gives the known thermal contribution. We show only the result because it can be calculated following the same steps done for the correction which we will show later. The main aspect is that in the free case (non-random Dirac equation) the probability of excitation, i.e., $E > 0$ (modulo the selectivity) of the Unruh-DeWitt detector presents a Bose-Einstein factor \cite{langlois}. We get $lim_{\tau\rightarrow\infty}R_{0}(E,\tau) = W_{0}(E,\tau)$ where

\begin{equation}\label{freefermidirac}
W_0(E,\tau) = \frac{1}{64\pi} \frac{E\tau}{e^{2\pi E/ \alpha}-1 } (4\alpha^4 + 5E^2\alpha^2 + E^4).
\end{equation}

Note that the second and third terms in Eq. (\ref{eq:trace-exp}) are the main corrections up to $\sigma^2$ order, see Eqs. (\ref{freecase}) and (\ref{eq:correction_S}), and the last term is of the order $\sigma^4$. So, taking the limit $\epsilon \rightarrow 0$, the leading contribution due to horizon events fluctuations is given by:

\begin{equation}
R_1(E,\tau)=-\frac{8i}{2\pi^2}\left(\frac{\alpha}{2}\right)^9\int_{-\tau}^{\tau}d\xi e^{-iE\xi}\int_{|\xi|}^{2\tau-|\xi|}d\eta
\left\{
\frac{-\sigma^{'2}_1(\eta)+\sigma^{'2}_2(\eta)}{\sinh^9{(\frac{\alpha\xi}{2})}}
\right\},
 \label{eq:lead-correction}
\end{equation}
where we defined $\sigma^{'2}_1(\eta) = {\sigma}^2_1(\eta)\cosh{(\frac{\alpha\eta}{2})}$ and $\sigma^{'2}_2(\eta) = {\sigma}^2_2(\eta)\sinh{(\frac{\alpha\eta}{2})}$. The integration over $\eta$ it's easier to be done using algebraic computational programs.
Let us write these integrations in terms of two functions $f_1(\tau,\xi)$ and $f_2(\tau,\xi)$. Now we can obtain the expression:

\begin{equation}
R_1(E,\tau)=\frac{-8i}{2\pi^2}\left(\frac{\alpha}{2}\right)^9\int_{-\tau}^{\tau}d\xi e^{-iE\xi}
\left\{
\frac{f_1(\tau,\xi)+f_2(\tau,\xi)}{\sinh^9{(\frac{\alpha\xi}{2})}}
\right\}.
 \label{eq:lead2-correction}
\end{equation}
To perform these integrals it is convenient to replace $\chi=\alpha\xi$ and express each integral as

\begin{align}
\nonumber\frac{1}{\alpha}\int_{-\alpha\tau}^{\alpha\tau}d\chi  e^{-i\frac{E}{\alpha}\chi}h(\tau,\chi)&=
\frac{1}{\alpha}\int_{-\infty}^{\infty}d\chi  e^{-i\frac{E}{\alpha}\chi}h(\tau,\chi) \\
&- \frac{1}{\alpha}\left[\int_{-\infty}^{-\alpha\tau}d\chi  e^{-i\frac{E}{\alpha}\chi}h(\tau,\chi)
+ \int_{\alpha\tau}^{\infty}d\chi  e^{-i\frac{E}{\alpha}\chi}h(\tau,\chi)\right],
 \label{eq:finite-int}
\end{align}
where $h(\tau,\chi)$ are the terms in the curly brackets from Eq. (\ref{eq:lead2-correction}). Replacing $\chi\rightarrow -\chi$ in the first integral into squared brackets, we can simplify all brackets to a single integral given by:

\begin{align}
K(E,\tau)=-\frac{16}{\alpha}\left(\frac{\alpha}{2}\right)^9
\int_{\alpha\tau}^{\infty}d\chi\sin{\left(\frac{E\chi}{\alpha}\right)}h(\tau,\chi).
 \label{eq:vanish-int}
\end{align}
The nature of this term is associated with the switching on and off of the interaction between the detector and the background field.
It can be neglected if we assume large proper time intervals. However, for short proper times intervals it must be taken into account.

We need now to evaluate the infinity range integrals in Eq. (\ref{eq:finite-int}). These integrals are given by:

\begin{figure}
\begin{center}
\includegraphics[width=0.4\textwidth]{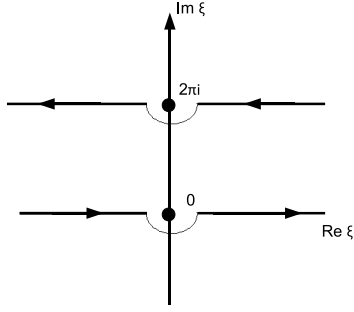}
\end{center}
\caption{Contours of integration to Eq. (\ref{eq:special-int}) }\label{fig1}
\end{figure}

\begin{equation}
I_n(E,\tau) = \frac{1}{\alpha}\int_{-\infty}^{\infty}d\chi\frac{e^{-i\frac{E \chi}{\alpha}}}{\sinh^n(\chi/2)}h(\tau,\chi).
\label{eq:special-int}
\end{equation}
In these cases $n = 9$. These integrals can be evaluated choosing the contours given by Fig. (\ref{fig1}) and using the residue theorem done in Refs. \cite{akms11, det1}. However, note that we have odd order poles. In fact, these integrals will be associated with a nine order residue of
$h(\chi)$ at $\chi = 0$. So, performing such integral the correction to the response function for the Unruh-DeWitt detector is:

\begin{align}\label{eqw784}
W_1(E,\tau)=\frac{ \sigma^2(\tau) E^2 }{e^{2\pi E/\alpha}+1 }\times F(E, \alpha)
\end{align}
where

\begin{equation}
F(E, \alpha) = \frac{\left(-298262 \alpha^6-29302 \alpha^4 E^2+12817 \alpha^2 E^4+1842 E^6\right)}{1260
   \pi\alpha}
\end{equation}
and now $\sigma^2(\tau) \approx \sigma^2\sinh(\alpha\tau)\approx \sigma^2\alpha\tau$, when $\alpha\tau \ll 1$.

We note that the correction to the response function shows a Fermi-Dirac factor instead of a Bose-Einstein factor found in the free case. If a temperature could be defined here, it would be given by $T = 2\pi/\alpha$. However there is some remarks to discussed here. Remember that in the Minkowski reference frame the noise function $\mu({\bf x})$ depends only on the spatial coordinates, while in the Rindler system the noise function depends on the proper time $\tau$, in other words the noise is a function of $\xi$ and $\eta$, and now we have a non-equilibrium behavior of the field measured by the detector. In this case we can define a local temperature, see  Refs. \cite{Ortiz} and \cite{kp10}, which depends on the proper time by the function $\mu(\xi,\eta)$. It was showed that the temperature which appears in the response function correction is an averaged temperature over the fluctuations and the global temperature is related  with the thermal equilibrium.

The last result shows that the lack of stationarity in the accelerated frame does not necessarily implies that a thermal-like response function with
local temperature is forbidden. With such above considerations a possible interpretation of our result is that light-cone fluctuations in this model
lead to the emergence of a local temperature as seen by accelerated observers.

So far we have evaluated the horizon fluctuating contribution to the response function for an Unruh-DeWitt detector coupled to a massless fermionic field. Instead using a detector like that one considered in Eq. (\ref{eq:lagrangian}), it is also possible to use another kind of detectors. It is known that the interaction Hamiltonian for the scalar field coupled with the Unruh-DeWitt detector is a monopole-moment \cite{unruhdet}. In other words, it is a two level system operator coupled with a scalar field, hence it is linear in the field. It gives a response function which depends on the two point instead of the four point Green function found to the fermionic field case. Thus, instead of Hamiltonian given by Eq. (\ref{eq:lagrangian}), we can find an analogue kind of monopole-moment detector similar to the scalar case Hamiltonian.  Despite the fact that this detector is less realistic, one can consider it equipped with a spinor $\Theta$ and coupled linearly to the field via the Hamiltonian:

\begin{equation}
 \mathcal{H}_{int}=\overline{\Theta}(\tau)\Psi(x(\tau))+\overline{\Psi}(x(\tau))\Theta(\tau),
\label{eq:lagrangian11}
\end{equation}
where $x^{\mu}(\tau)$ is again the world line of the two level system with respect to the proper time $\tau$, and $\Theta(\tau)$ is the fermionic
monopole moment operator. To keep the Hamiltonian stationary to the observer's proper reference frame it is also necessary to  Fermi-Walker transport the monopole operator and the Dirac field, in same way that was done above using Eqs. (\ref{eq:FW-fields})-(\ref{eq:FW-fields3}). One can also formulate a perturbation theory in the interaction picture where it is assumed that $\hat{\Theta} = S_{\tau}\Theta$ and $\bar{\hat{\Theta}} = \bar{\Theta}S_{\tau}$ rather than $\Theta$ and $\bar{\Theta}$ obey the time-evolution equations.

Following these steps one can find the new response function:

\begin{eqnarray}
 \mathcal{F}(E,\tau,0)= \frac{1}{4}Tr\left[\gamma^0\int_{0}^{\tau}d\tau'\int_{0}^{\tau}d\tau'' e^{-iE(\tau'-\tau'')}
 \right. \\ \nonumber \left. \times \langle 0|\hat{\Psi}_{\alpha}(x(\tau'))\overline{\hat{\Psi}}_{\beta}(x(\tau''))|0\rangle\right].
\label{eq:response-function}
\end{eqnarray}
This result is similar to that one found in Ref. \cite{t99}.

The response function is proportional to the two point Wightman function instead of the four point function found previously. The calculation is done in a finite time interval as was done in Eqs. (\ref{eq:finite-int}, \ref{eq:vanish-int}), and up to first order in $\tau$ we have that the correction due to horizon fluctuations is

\begin{equation}\label{eq:response-fermion22}
W_1(E,\tau)\approx \frac{\sigma^2(\tau) \left(-8059 \alpha^4E - 55\alpha^2E^3 + 3684E^5\right)}{60 \pi\alpha(e^{2\pi E/\alpha}-1) }
\end{equation}
the $\sigma^2(\tau)$ is the same as in Eq. (\ref{eqw784}). This integral was evaluated using Eq. (\ref{eq:special-int}) with $n = 6$ and we also considered the same contour given in Fig. (\ref{fig1}). 

In the case without fluctuations the two point function is given by Eq. (\ref{freecase}), and the response is

\begin{equation}
W_0(E,\tau) = \frac{\pi\tau [E^2 + (\alpha/2)^2]} {e^{\frac{2\pi E}{\alpha}}+1},
\label{eq:response-fermion2}
\end{equation}

Comparing Eqs. (\ref{eq:response-fermion22}) and (\ref{eq:response-fermion2}) we note that the event horizon fluctuation modifies the Fermi-Dirac response found in the free case. It appears a Bose-Einstein factor. As we discussed, it is also possible to define a local temperature, $T = 2\pi/\alpha$, in this non-equilibrium situation and, in a similar way, the effective noise strength is proportional to the proper time.

\section{Conclusions}\label{section5}

Recently it was proposed in condensed matter physics an analog model for quantum gravity effects \cite{kms10} and in Ref. \cite{pe21} the authors extended this theory to a real massive scalar field. In Ref. \cite{akms11} it was considered a massless scalar field near a four-dimensional Schwarzschild black hole with event horizon fluctuations. They used a perturbation theory similar to that one previously considered in fluctuating disordered media.

In this paper we have used these ideas to study the behavior of a massless fermionic field near a fluctuating black hole horizon using the techniques described before. First, we developed a perturbation theory associated to a massless fermionic field in the presence of random noise. After performing the calculations we have applied these results to investigate the thermal radiation near a fluctuating horizon. We showed that horizon fluctuation implies that the Unruh-DeWitt detector which has a Bose-Einstein factor in the response function associated with the free case has now a correction which has Fermi-Dirac factor. We would like to stress that we are assuming that it is possible to define a local temperature in this non-equilibrium situation. In a less realistic detector, coupled linearly with a Dirac field, the correction has also an apparent change in the distribution measured by in the response function with unchanged temperature. In this situation the leading term (case without fluctuations) has a Fermi-Dirac factor and in the correction appears a Bose-Einstein factor.

Since the de Sitter horizon has also a thermal spectrum measured by the Unruh-DeWitt detector \cite{gp04} for both, bosons and fermions, a natural extension of our work is to study the effects of a fluctuating de Sitter horizon to such kind of coupling. It is also natural to enquire the effects of introducing a time dependent randomness \cite{s88} in place of the Eqs. (\ref{noise1}) and (\ref{noise2}). Such random coefficients allows the possibility of particle production from the beginning \cite{pe11}. These subjects are under investigation by the authors.

\appendix
\renewcommand\thesection{Appendix \Alph{section}}

\section{One-loop correction to positive frequency Wightman function}

Let us begin from Eq. \eqref{eqs12} which gives the correction to the positive frequency Wightman function. The self energy is given in Eq. \eqref{self1}. By dimensional regularization, only the first term do contribute which corresponds to $2\pi i \omega^2\gamma^0$. Replacing this result in Eq. \eqref{eqs12} this expression can be separated in three different parts $\gamma^{\mu\dagger}\gamma^{\delta}k_\mu k_\delta = \delta_{ij}\omega^2 - 2\omega k_j\gamma^j\gamma^0 - \gamma^i\gamma^jk_ik_j$ writing Eq. \eqref{eq:correction-coor} as

\begin{eqnarray}
\bar{S}^{(1)}(t,{\bf x}) = 2\pi^2i\sigma^2\gamma^0\int\frac{d^3{\bf k}}{(2\pi)^3}
\left\{I_6 + 2k_j\gamma^j\gamma^0I_5 - \gamma^i\gamma^jk_ik_jI_4\right\}e^{-i {\bf k}\Delta {\bf x}},
\label{eq:WF-correction}
\end{eqnarray}
where $I_n$ are integrals defined by:

\begin{equation}
I_n = \int\frac{d\omega}{(2\pi)}\frac{\omega^n}{(\omega^2 - {\bf k}^2)^2}e^{-i\omega(t - t')}.
\label{eq:integral-In}
\end{equation}

To calculate the positive Wightman function, we must evaluate Eq. \eqref{eq:integral-In} over the contour of integration for the variable $\omega$ as is defined in Ref. \cite{pe1}. In this case only the pole $\omega=|{\bf k}|$ is chosen and the residues theorem is used. It yields to

\begin{equation}
I_n = \frac{i}{4}\omega^{n-3}[(n-1) - i|{\bf k}|\Delta t]e^{-i{\bf |k|}\Delta t} .
\end{equation}

The expression above is simplified and can be written in terms of the momenta as

\begin{align}
 \nonumber \bar{S}^{(1)}(t,{\bf x})&= 2\pi^2i\sigma^2\gamma^0\int\frac{d^3{\bf k}}{(2\pi)^3} \left\{\frac{i{\bf |k|}^3}{4}\left[-i\Delta t{\bf |k|} + 5\right]
 \right. \\ \nonumber &+ 2ik_i\gamma^i\gamma^0\frac{{\bf |k|}^2}{4}\left[-i\Delta t{\bf |k|} + 4\right]  \\
 &\left. -  i\gamma^i\gamma^jk_ik_j\frac{{\bf |k|}}{4}\left[-i\Delta t{\bf |k|} + 3\right]\right\}e^{-i{\bf |k|}\Delta t + i{\bf k}.{\bf \Delta x}}.
\end{align}

The last two integrals can be simplified using the fact that, $\int d^3{\bf k}k_i\gamma^if(|{\bf k}|)e^{-ik(x - x')} = \gamma^i\partial_i\int d^3{\bf k}f(|{\bf k}|)e^{-ik(x - x')}$. So our integrals are from the type

\begin{equation}
B_1^{(n)} = \int\frac{d^3{\bf k}}{(2\pi)^3}|{\bf k}|^ne^{-ik(x - x')},
\end{equation}
the expression is simplified by

\begin{align}
\nonumber\bar{S}^{(1)}(t,{\bf x}) = &2\pi^2i\sigma^2\gamma^0\left(\frac{5i}{4}B_1^{(3)} + \frac{1}{4}\Delta tB_1^{(4)} - 2\gamma^j\gamma^0\partial_jB_1^{(2)}   \right. \\ &\left. +
\frac{i}{2}\gamma^j\gamma^0\Delta t\partial_jB_1^{(3)} + \frac{3i}{4}\gamma^i\gamma^j\partial_i\partial_jB_1^{(1)}
 + \frac{\Delta t}{4}\gamma^i\gamma^j\partial_i\partial_jB_1^{(2)} \right).
\label{eq:Correction2}
\end{align}

After calculating the angular parts from the above integrals it is useful to define a function $g_n(x, x')$ such that the function $B_1^{(n)}$ and its
derivatives can be expressed in terms of $g_n(x, x')$ by the following expression

\begin{equation}
 g_n(x, x') = \int_0^\infty dk k^n(e^{-ik(\Delta t - \Delta{|{\bf x}|})} - e^{-ik(\Delta t + \Delta{|{\bf x}|})}).
\end{equation}
In the last expression $k$ is a real variable not a four-vector, thus, it can be put out from the integral through derivatives with respect to $\Delta t$ writing the last equation as:

\begin{equation}
 g_n(x, x') = \frac{1}{(-i)^n}\frac{\partial^n}{\partial\Delta t^n}\int_0^\infty dk (e^{-ik(\Delta t - \Delta{|{\bf x}|})} - e^{-ik(\Delta t + \Delta{|{\bf x}|})}).
\end{equation}
This expression is related to the Wightman positive function for the scalar field, in such a way that the $g_n(x, x')$ function may be written by (see Ref. \cite{pe1})

\begin{equation}
 g_n(x, x') = \frac{1}{(-i)^n}\frac{\partial^n}{\partial\Delta t^n}\left(\frac{-2i|\Delta {\bf x}|}{(\Delta t - i\epsilon)^2 - |\Delta{\bf x}|^2}\right).
\label{eq:g-escalar}
\end{equation}

Thus $B_1^{(n)}$'s integrals can be simplified by the expression

\begin{equation}
 B_1^{(n)} = \frac{1}{i(2\pi)^2|\Delta {\bf x}|}g_{n + 1}.
\label{eq:Bn}
\end{equation}

Finally, substituting  Eqs. (\ref{eq:g-escalar}) and (\ref{eq:Bn}) into Eq.\eqref{eq:Correction2} we get the first contribution to the Wightman function which is given by Eq.\eqref{eqfim1}.

\section{One-loop correction to the Rindler noise}\label{apendiceb}

With the Rindler coordinates defined in Eqs. (\ref{eq: Rindler-coordiantes}) we are able to define the Rindler noise for the Dirac field by: $g_{1/2}(\tau) = S_\tau\langle S^{(1)}(x(\tau), x(\tau'))\rangle S_{-\tau'}$. So using this definition together to Eqs. (\ref{eqfim1}), (\ref{eqfim2}) and (\ref{eqfim3}) we find that in accelerated reference frame

\begin{eqnarray}\label{eqg12}
 \mathcal{S}^{(1)}_{1/2}(\xi, \eta) = \frac{\sigma^2}{2D^6}\gamma^0 S_{-\eta}\left(\frac{2}{\alpha}\right)^5 \sinh^5\frac{\alpha\xi}{2}
\left[\frac{2}{\alpha}\sinh\frac{\alpha\xi}{2}F(z) + i\epsilon G(z)\right] \\ \nonumber
+ \frac{\sigma^2}{2D^6}\gamma^1 S_{-\eta}\left(\frac{2}{\alpha}\right)^5\sinh^5\frac{\alpha\xi}{2}
\left[\frac{2}{\alpha}\sinh\frac{\alpha\xi}{2}H(z) + i\epsilon J(z)\right],
\end{eqnarray}
where we have defined $z=\frac{\alpha\eta}{2}$. The functions $F(z), G(z), H(z)$ and $J(z)$ are defined by

\begin{align}\label{eq:fghj}
 F(z) &= \cosh z\left(-3684\cosh^5 z + 7539\cosh^3 z\sinh^2 z  \right. \nonumber\\
         &\left.- 8328\cosh z\sinh^4 z - 1005\cosh^{-1}z\sinh^6 z\right),
\end{align}
\begin{align}\label{eq:fghj2}
 G(z) &= \cosh z\left(18504\cosh^4 z - 24732\cosh^2 z\sinh^2 z \right. \nonumber\\
         &\left. + 9354\sinh^4 z \right),
\end{align}
\begin{align}\label{eq:fghj3}
 H(z) &= \cosh z\left(-1660\cosh^4 z\sinh z - 2024\cosh^2 z\sinh^3 z \right. \nonumber\\
         &\left. - 156\sinh^5 z \right),
\end{align}
\begin{align}\label{eq:fghj4}
 J(z) &= \cosh z\left(6620\cosh^3 z\sinh z + 5056\cosh z\sinh^3 z \right. \nonumber\\
         &\left.  + 12\sinh^5 z\cosh^{-1} z \right),
\end{align}
expanding the denominator in Eq. (\ref{eqg12}) and after some calculations we find a simplified expression for the correction of the Rindler's noise given in Eq.\eqref{eq:correction_S}.

\bigskip
\centerline{\bf Acknowledgments}
The authors would like to thanks E. Arias and G. Menezes for helpful discussions. This work is partially supported by the Conselho Nacional de
Desenvolvimento Cient\'{\i}fico e Tecnol\'{o}gico - CNPq and Coordena\c{c}\~ao de Aperfei\c{c}oamento de Pessoal de N\'{\i}vel Superior - CAPES (Brazilian Research Agencies).

\end{document}